\documentclass[journal]{IEEEtran}
\makeatletter
\def\footnoterule{\kern-3\p@
  \hrule \@width 2in \kern 2.6\p@} % the \hrule is .4pt high
\makeatother

\usepackage{tikz}
\usetikzlibrary{shapes, arrows.meta, positioning}

\usepackage{graphicx}
\usepackage{amsmath}
\usepackage{url}
\usepackage{authblk}
\usepackage{graphicx}
\usepackage{cite}
\usepackage{algorithm}
\usepackage{algpseudocode}
\usepackage{multirow}
\usepackage{array}
\usepackage[dvipsnames]{xcolor}
\usepackage{authblk}
\usepackage{enumitem}
\author[1]{Muhammad Ali Hassan Ahmad}
\author[2]{Muhammad Hashim Ali}
\author[3]{Muhammad Ali Amer}
\author[4]{Muhammad Naiman Jalil}
\author[5]{Muhammad Hassan}
\author[6]{Affan Rauf}

\affil[1]{Lakehead University, Ontario, Canada}
\affil[2,3,6]{National University of Computer and Emerging Sciences, Pakistan}
\affil[4]{United Arab Emirates University, United Arab Emirates}
\affil[5]{German Research Center for Artificial Intelligence, Germany}

\begin{document}
\title{Enhancing Reliability of Symbolic Execution Tools for Smart Contract Analysis through Rule-Based False Positive Reduction}
\maketitle

\begin{abstract}
A blockchain is a decentralized, secure ledger system that enables transparent and immutable record-keeping, essential for trust and security in digital transactions. Smart contracts are self-executing agreements encoded on a blockchain, enabling different parties to fulfill the terms of the agreement automatically. These contracts trigger corresponding actions when conditions are met, ensuring decentralized and transparent transactions.  Writing reliable smart contracts is challenging due to the lack of standardization. To find security vulnerabilities, tools based on various approaches, including symbolic execution, are used.  However, these tools often report a large number of false positives, raising concerns about their reliability.  The time and effort spent investigating false positives diverts resources from addressing actual vulnerabilities. Therefore, such tools must also be evaluated according to the rate of false positives they exhibit.  More importantly, the algorithms and heuristics used by the tools must be enhanced to distinguish between true vulnerabilities and false alarms. 
In this paper, we first demonstrate the prevalence of false positives in vulnerability reports generated by Mythril, a symbolic execution-based analysis tool for Ethereum smart contracts. We analyze the root causes of these inaccuracies and devise a rule-based approach based on the gained insight to reduce false positives.  We implement our rules for the most impactful vulnerabilities in Mythril and assess the effectiveness of our approach. Our results show a significant  reduction in false positives without compromising the detection of true vulnerabilities, thus enhancing the tool's reliability.
\end{abstract}

\setlength{\parindent}{1.5em}
\setlength{\parskip}{0pt}

\section{Introduction}
\label{sec: Introduction}
Blockchains have gained popularity due to their convenience of secure, transparent, and decentralized record-keeping, which eliminates intermediaries and lowers fraud risk. A blockchain is a decentralized distributed ledger system that records transactions across multiple systems, once a transactions is recorded there appears no possibility for later modification. Smart contracts are the brain that makes the blockchain function beyond a ledger. They are self-executing programs stored on the blockchain that automatically run and carry out the terms of an agreement when predefined conditions are met. They are used for secure and transparent management of transactions and record-keeping in blockchain systems. Once deployed on the blockchain, they also become permanent, requiring them to be thoroughly reviewed and tested before deployment to eliminate errors and vulnerabilities. However, writing and testing secure smart contracts can be challenging due to the lack of standardized practices and tools. Security attacks targeting vulnerable smart contracts have been on the rise, which has led to financial loss, erosion of trust, and other security issues. 

Several smart contract analysis tools exist that detect different vulnerabilities; however, most of these tools, even the most popular ones, report a large number of false positives along with real vulnerabilities. These large numbers of false positives are confirmed by multiple smart contract auditing bodies in the form of benchmarks. A high rate of false positives in a security analysis tool can be problematic for several reasons. For example, investing excessive time in analyzing reported false positives impacts developers' productivity and consequently may incur significant financial costs. For auditors, false positives result in a shift of emphasis from critical issues and, as a result, may pose a risk to the reliability of smart contracts. As end users rely on the integrity of smart contracts, which are validated using various tools, any challenge to the reliability of these tools also raises concerns about the efficacy of the auditing process, affecting the trust users have in the functionality of these contracts.

The tools must accurately detect vulnerabilities rather than producing a high number of false positives. Previous research addresses either reporting smart contract vulnerabilities or reporting other problems, including false negatives, in the analysis tools. To the best of our knowledge, no previous work has focused on reducing false positives in symbolic analysis tools.

This paper highlights false positives in smart contract analysis tools as a threat to the reliability of the whole smart contract ecosystem. It proposes a rule-based approach to minimize false positives associated with six significant vulnerabilities in symbolic execution based tools, implements these rules in Mythril's \cite{a16} execution, and evaluates their effectiveness on Gigahorse\cite{a15} smart contracts, showing that Mythril, augmented by our rules, considerably reduces the number of false positives. 

\section{Related Work}
\label{sec:literature_review}
Asem Ghaleb et al.\cite{a1} propose a systematic approach to evaluate the proposed analysis tools and their effectiveness. To do this, the author uses a tool called \cite{a19} SolidiFI. SolidiFI injects bugs into all potential locations in a smart contract to introduce targeted security vulnerabilities. SolidiFI then checks the generated buggy contract using the static analysis tools and identifies the bugs that the tools cannot detect, along with identifying the bugs reported as false positives. Paper has a dataset of 50 smart contracts in which they inject 9369 distinct bugs. These bugs were evaluated by 6 smart contract analysis tools, which were Oyente \cite{a20},  Securify \cite{a21}, Mythril \cite{a16}, SmartCheck \cite{a7},  Manticore \cite{a17}, and Slither \cite{a24}. The experiment injects the bugs into the smart contract and then evaluates them using the analysis tools, and then evaluates whether the tool correctly identified all the injected bugs. This Paper is mainly focused on false negatives. Many of the detected bugs were not caught by the analysis tools such as Mythril, which failed to detect the largest set of bugs in the experiments.

Reza M. Parizi et al.\cite{a2} mainly try to answer two different questions relating to smart contract analysis tools regarding effectiveness and accuracy. There are 4 main tools that \cite{a2} uses, namely, Oyente, Mythril,  Securify, and SmartCheck. The paper uses 10 smart contracts to experiment on. The experiment selects a new tool randomly and a new smart contract randomly. It then applies the tool to the smart contract and collects the results. \cite{a2} showed that the SmartCheck tool is statistically more effective than the other automated security testing tools at a 95 percent significance level. Concerning accuracy, Mythril was found to be significantly accurate, issuing the lowest number of false alarms among peer tools.  On evaluating the accuracy, the Mythril tool showed the highest accuracy score; however, the Oyente tool again performed the worst in accuracy.

Ardit Dika et al.\cite{a3} examine the different audited and vulnerable smart contracts and use different analysis tools to gauge their effectiveness, accuracy, and consistency. \cite{a3} analyses both bytecode and solidity contracts. It focuses on both false negatives and false positives. \cite{a3} uses only 21 audited smart contracts and 24 vulnerable smart contracts. 4 tools were used to analyze the smart contracts: Oyente,  Securify, Remix  \cite{a25}, and SmartCheck  \cite{a7}. \cite{a3} analyzes different smart contracts and analyzes them according to the false positives and false negatives. It then creates a vulnerability matrix to find which tool accurately catches what vulnerability. According to \cite{a3}, SmartCheck found the most vulnerabilities.

Alexander Mense et al.\cite{a4} summarize known vulnerabilities in smart contracts found by literature research. It compares currently available contract analysis tools for their capabilities to identify and detect vulnerabilities in smart contracts based on a taxonomy for vulnerabilities. \cite{a4} lists different vulnerabilities present in smart contracts and classifies them according to the severity level. These vulnerabilities include, but are not limited to, gas-less send, external calls, and mishandled exceptions. \cite{a4} then uses analysis tools, namely, Oyente, Securify, Remix, SmartCheck, F* \cite{a26}, Mythril, and Gasper \cite{a30}. Re-entrancy ranks the highest among all the vulnerabilities, with 6 out of 7 tools detecting it. The results show the different analysis tools detecting different vulnerabilities.

S. Kim et al.\cite{a5} collected 391 papers and extracted 67 relevant to the smart contract analysis. These papers were classified into 3 different topics: static analysis for vulnerability detection, static analysis for program correctness, and dynamic analysis, which were then explored with further classifications. The selected papers focused on analyzing smart contracts using either static or dynamic techniques. In the context of static analysis for vulnerability detection, symbolic execution is employed to simulate concrete executions with inputs represented as symbolic values. Additionally, concolic testing emerges as the dominant technique for detecting vulnerabilities in Ethereum Virtual Machine (EVM) bytecode. \cite{a5} evaluates that dynamic analysis primarily relies on techniques such as fuzzing and runtime verification to analyze program behavior during execution.

Enmei Lai et al.\cite{a6} analyze the integer overflow of Solidity smart contracts. It summarizes 11 kinds of integer overflow features. After this analysis, a static detection tool for integer overflow is designed based on the idea of the SmartCheck tool \footnote{https://github.com/smartdec/smartcheck}. The designed tool can detect integer overflow vulnerabilities in smart contracts. 7000 smart contracts were tested on the designed tool, out of which 430 smart contracts had an integer overflow vulnerability. \cite{a6} evaluates 7000 smart contracts using the designed tool, and the accuracy of the detection results is almost 100 percent. There were 109 smart contracts with multiplication overflow vulnerabilities, 218 smart contracts with addition overflow vulnerabilities, and 103 with subtraction overflow vulnerabilities. \cite{a6} also compares their tool with other smart contract vulnerability detection tools.

Tikhomirov et al. \cite{a7}. Provide a taxonomy of Solidity issues and introduce SmartCheck, an extensible static analyzer (Java). Evaluated on real-world contracts against manual audits, SmartCheck reports a broad range of security findings with comparatively high accuracy and low false-positive rates.

% Sergei Tikhomirov et al.\cite{a7} provide a comprehensive classification of code issues in Solidity and also implement SmartCheck, which is an extensible analysis tool that detects issues. \cite{a7}  is evaluated on a big dataset of real-world contracts, and then they are compared with manual audits. This allows for testing a wide range of input values and uncovering hidden vulnerabilities that might not be detected by other static analysis tools. The paper compiles a list of 4 main categories of issues, which include functional issues, security issues, operational issues, and developmental issues. There are various issues in each category. A tool called SmartCheck is implemented in Java. Several issues are identified, and these are checked with three available tools, Oyente, Remix, and  Securify, on three contracts. The results show that SmartCheck was able to detect a variety of security vulnerabilities with high accuracy and low false positive rates, 99.9 percent of contracts have issues, and 63.2 percent of contracts have critical vulnerabilities.

Jiaming Ye et al.\cite{a8} present Clairvoyance \footnote{https://github.com/nikitastupin/clairvoyance}, a cross-function and cross-contract static analysis by identifying infeasible paths to detect re-entrancy vulnerabilities in a smart contract. It presents a large-scale empirical study to evaluate the effectiveness of three recent general-purpose static tools using 11714 real-world contracts. To address the re-entrancy issue,\cite{a8} developed Clairvoyance, a static analysis tool that uses interprocedural control-flow analysis to analyze the code of smart contracts and identify potential vulnerabilities. According to the findings, the tools are not sufficiently effective in handling the re-entrancy bug. The reason is that most of their reports are false negatives. To avoid these false positives happening to the tool, \cite{a8} audits incorrect reports manually. For the number of detection results of the four static tools, Slither reports 162 vulnerabilities in total, of which 3 reports are true positives, while the other 159 reports are false positives. Oyente has at least 28 reports, of which 4 results are true positives, while the rest 24 reports are false positives.

Sefa Akca et al.\cite{a9} propose techniques that will allow complete automated analysis of smart contracts, using both static and dynamic techniques. The author injects well-known vulnerabilities into smart contracts. Four major works in \cite{a9} are static checks, test generators, run-time monitoring, fault seeding tools, and empirical evaluation. The approach for analysis of smart contracts includes three main components: 1) A vulnerability detection technique, SolAnalyser \cite{a9}. 2) An automated input generation for smart contracts, 3) Use of a tool, MuContract. SolAnalyser, with static and dynamic checks supported by test generation, was effective at detecting vulnerabilities across all 1838 contracts and the 12866 mutated versions, with a precision of 72 percent and a recall rate of 100 percent. The technique was capable of detecting more types of vulnerabilities than all five existing analysis tools used in the experiment.

Yuchiro Chinen et al.\cite{a10} present a static analysis tool named RA (Re-entrancy analyzer) to analyze smart contract vulnerabilities against re-entrancy attacks. \cite{a10} aims to design an inter-contract static analysis tool that uses only EVM bytecodes as input, eliminates false negatives and false positives, and does not require analysts to have prior knowledge of the attacks on contracts. We can use \cite{a10} to evaluate and analyze smart contract vulnerabilities, especially re-entrancy issues, in our research. \cite{a10} Develops a static analysis tool that focuses on inter-contract analysis using EVM bytecodes as input, effectively reducing false negatives and false positives. The novelty of the RA tool \cite{a10} lies in its ability to perform analysis without requiring analysts to possess prior knowledge of specific attacks on contracts.

Joao F. Ferreira et al.\cite{a11} present SmartBugs\footnote{https://github.com/smartbugs/smartbugs}, an extensible and easy-to-use execution framework that simplifies the execution of analysis tools on smart contracts written in Solidity, the primary language used in Ethereum. 
One of the primary objectives of SmartBugs \cite{a11} is to enhance the reproducibility of research focused on automated reasoning and testing of smart contracts. \cite{a11} demonstrates the ease of integrating new tools into SmartBugs and compares its performance with existing tools. Specifically, \cite{a11} extended SmartCheck and utilized SmartBugs to showcase the significant improvement in vulnerability detection, particularly related to Bad Randomness, Time Manipulation, and Access Control.

Xingrun Yan et al.\cite{a12} propose a semantic analysis-based method for the smart contract code to effectively defend against vulnerability attacks and improve detection efficiency. \cite{a12} implements the approach as a self-contained tool, which is then evaluated with 99 smart contracts and compared with other detection tools. The results demonstrate that the method can enhance both the accuracy and the ability to identify vulnerabilities. \cite{a12} used four datasets, each comprising distinct vulnerabilities, to measure the effectiveness of their method. The methodology accurately detected all instances of fallback function re-entrancy vulnerabilities and cross-function re-entrancy vulnerabilities across the datasets. The authors compared their methodology with two tools, Mythril and Slither \cite{a24}, which both failed to identify all instances of these vulnerabilities.

Nikolay Ivanov et al. \cite{a31} conduct an evolutionary analysis and classify 133 threat mitigation solutions for smart contracts using a five-dimensional taxonomy. They combine eight mitigation methods and map vulnerabilities addressed by these solutions. Their methodology includes 1. Classify solutions using taxonomy. 2. Synthesize the workflow of eight core methods. 3. Map vulnerability. 4. Analyze trends and gaps. 

The paper Daojing He et al. \cite{a32} systematically categorizes smart contract vulnerability detection tools into six classes. The paper evaluates 27 tools for their accuracy and multi-version handling capability. The results demonstrate that most tools focus on older smart contract versions, while the new tools involving machine learning are more accurate; however, they flag fewer vulnerabilities. The paper proposes to combine static, dynamic, and neural network-based tools for accurate multi-version detection. While the key limitations include the time-intensive nature of static methods, and the challenges dynamic tools face in detecting sequence-specific vulnerabilities, with few tools supporting multi-version analysis.

The paper Chong Chen et al. \cite{a33} systematically reviews the popular large language model, ChatGPT\footnote{https://platform.openai.com/docs/api-reference/introduction}, to assess its effectiveness and limitations in identifying smart contract vulnerabilities and to compare it to native tools. ChatGPT effectively identifies vulnerabilities like denial-of-service, front-running, and unchecked return values. However, it struggles with reentrancy and access control. The key limitations include the limited number of tokens and uncertainty in the analysis, which lead to erroneous or biased judgments. It also tends to produce false positives and relies heavily on specific protection mechanisms.

Zhiyang Chen et al.\cite{a34} explore the use of dynamically inferred invariants to enhance smart contract security and introduce the Trace2Inv framework for generating these invariants from transaction histories. Individual invariant guards effectively block real-world exploits with low gas overhead, and combining multiple guards increases coverage while maintaining a low false positive rate. Limitations include the risk of missing bypass strategies, biased hack selection, the need for dynamic updates tied to variables like Oracle prices, and challenges in integrating these guards with other DeFi protocols.

Yuqiang Sun et al.\cite{a35} present GPTScan, a tool that integrates GPT with static analysis to identify logic vulnerabilities in smart contracts. It achieves a low false positive rate of 4.39\% for non-vulnerable contracts and over 90\% precision for token contracts, making it effective for large-scale scanning of on-chain contracts. GPTScan outperforms existing tools in detecting vulnerabilities within the Web3Bugs and DefiHacks datasets, with static confirmation reducing two-thirds of false positives while minimally affecting false negatives. However, its modifier filtering may lead to inaccuracies, the static analysis lacks path sensitivity and could benefit from symbolic execution, and the use of a temperature parameter of zero results in deterministic outputs that limit creativity.

\section{Proposed Approach}

\begin{table*}[t]
\caption{Comparative Performance of Vanilla vs. Improved Mythril on Gigahorse Benchmarks}
\label{tab:all-fp-results-combined}
\centering
\renewcommand{\arraystretch}{1.1}
\resizebox{\textwidth}{!}{
\begin{tabular}{|l||c|c||c|c|c|c|c|c|}
\hline
\multirow{3}{*}{\bfseries Vulnerability Type} & \multicolumn{2}{c||}{\bfseries GH Invulnerable ($N=40$)} & \multicolumn{6}{c|}{\bfseries GH Vulnerable ($N=100$)} \\
\cline{2-9} 
& \multicolumn{2}{c||}{\bfseries False Positives (FP)} & \multicolumn{2}{c|}{\bfseries False Positives (FP)} & \multicolumn{2}{c|}{\bfseries False Negatives (FN)} & \multicolumn{2}{c|}{\bfseries True Positives (TP)} \\
\cline{2-9}
& \bfseries FP (V) & \bfseries FP (I) & \bfseries FP (V) & \bfseries FP (I) & \bfseries FN (V) & \bfseries FN (I) & \bfseries TP (V) & \bfseries TP (I) \\
\hline
\hline
State Access after an External Call & 15 & 0 & 57 & 0 & 0 & 0 & 0 & 0 \\ 
Assertion Violation & 0 & 0 & 0 & 0 & 0 & 0 & 0 & 0 \\
Dependence on Predictable Env. Variable & 36 & 0 & 18 & 0 & 0 & 0 & 0 & 0 \\
External Call To User-Supplied Address & 8 & 1 & 25 & 2 & 0 & 0 & 0 & 0 \\
Integer Arithmetic: Overflow and Underflow & 34 & 50 & 11 & 12 & 4 & 3 & 12 & 13 \\
Multiple Calls in a Single Transaction & 9 & 0 & 19 & 0 & 0 & 0 & 0 & 0 \\
\hline
\textbf{Total Functions Flgged} & \textbf{102} & \textbf{51} & \textbf{130} & \textbf{14} & \textbf{4} & \textbf{3} & \textbf{12} & \textbf{13} \\
\hline
\multicolumn{9}{p{1.0\textwidth}}
{\footnotesize GH: GigaHorse. FP: False Positives. FN: False Negatives. TP: True Positives. V: Vanilla Mythril. I: Improved Mythril.}
\end{tabular}
}
\end{table*}

\label{sec:proposed_approach}
This paper aims to enhance the reliability of symbolic execution-based smart contract analysis tools by reducing the rate of false positives.
This reduction directly increases precision, which is defined as:
\[
\text{Precision} = \frac{\text{True Positives}}{\text{True Positives} + \text{False Positives}}
\]
Improving precision generally increases a tool's reliability to detect vulnerabilities. High precision indicates that a most identified vulnerabilities are true vulnerabilities, and the tool is less likely to flag false positives, making the tool's positive predictions more reliable.

\subsection{Approach: Rule-Based False Positive Reduction}
\label{sec:rule-based-approach}
We define vulnerability-specific rules to minimize false positives. These rules, directly patched in the implementation of a tool, help identify and block false positives within the symbolic execution process. These rules which often function as constraints are derived from characteristics of the analyzed smart contracts, vulnerabilities, and known patterns of false positives reported by analysis tools. Each rule is presented in Section~\ref{sec:tool-improvement}.

\subsection{Implementation and Rationale}
We implement our proposed rules in Mythril’s \cite{a16} implementation to assess the effectiveness of our approach. We selected Mythril due to its popularity, widespread adoption, symbolic execution-based nature, open-source nature, active maintenance, and capability to report a broad spectrum of significant vulnerabilities. For reproducibility, we utilized Mythril version v0.23.10 with commit \emph{21b3e92} of the Mythril's git repository\footnote{\url{https://github.com/ConsenSysDiligence/mythril/commit/21b3e92747f6b5ef1e22825756ba63f5b1c58621}}.

\subsection{Evaluation Methodology}
\label{sec:evaluation-methodology}

To assess and improve Mythril’s \cite{a16} precision, we utilized reputable sources of ground truth, specifically, Gigahorse Benchmarks \cite{a15}. These benchamarks provide open-source, labeled smart contracts, curated collections in both source and binary formats, each labeled with vulnerabilities. For reproducibility, we utilized commit \emph{ebcf377} of the Gigahorse Benchmarks repository\footnote{\url{https://github.com/nevillegrech/gigahorse-benchmarks/commit/ebcf37769d46253ae540d496cd84445120c57188}}.

Our evaluation employs a controlled, iterative comparison against distinct benchmark datasets, as summarized in Table \ref{tab:datasets}. The Gigahorse Benchmarks dataset, totaling 140 contracts, comprises two distinct collections critical for evaluation: 100 vulnerable contracts derived from the \texttt{vulnerable-sources} directory, each explicitly labeled with known vulnerabilities; and 40 invulnerable contracts from the \texttt{invulnerable-bytecode} directory, which are confirmed to be clean. This structure allows us to assess the tool's performance against both known issues and benign code. The Benchmark labels serve as a reference to determine whether a vulnerability flagged by Mythril corresponds to an actual issue. Any vulnerability flagged by Mythril but absent in the ground truth reference was considered a false positive.

The methodology is structured to quantify the improvement achieved by our rule-based approach:
\begin{enumerate}
\item \textbf{Baseline Assessment:} We apply the vanilla Mythril implementation to the datasets. The reported vulnerabilities are compared against the ground truth labels to establish a baseline false positive count.
\item \textbf{Enhanced Analysis:} The same process is repeated using our improved version of Mythril, which integrates the proposed rules for false positive reduction.
\end{enumerate}
This comparative analysis allows us to measure the absolute reduction in false positives and validate the practical benefits of integrating our approach into production-grade analysis tools.

\renewcommand\arraystretch{1}
\begin{table}[!t]
\caption{Datasets Used in This Study}\label{tab:datasets}
\resizebox{\columnwidth}{!}{%
\begin{tabular}{|c|c|}
\hline
\bfseries Dataset & \bfseries Number of Smart Contracts \\
\hline
Gigahorse Benchmarks (Vulnerable) & 100 \\
Gigahorse Benchmarks (Invulnerable) & 40 \\
\hline\hline
\textbf{Total} & 140 \\
\hline
\end{tabular}}
\end{table}

% \vspace*{-2\baselineskip}

\section{Baseline Evaluation}
\label{sec: Analysis and Benchmarking of Vulnerabilities}

\begin{figure}[t!]
\centering
\begin{minipage}{1\columnwidth}
  \includegraphics[width=\linewidth]{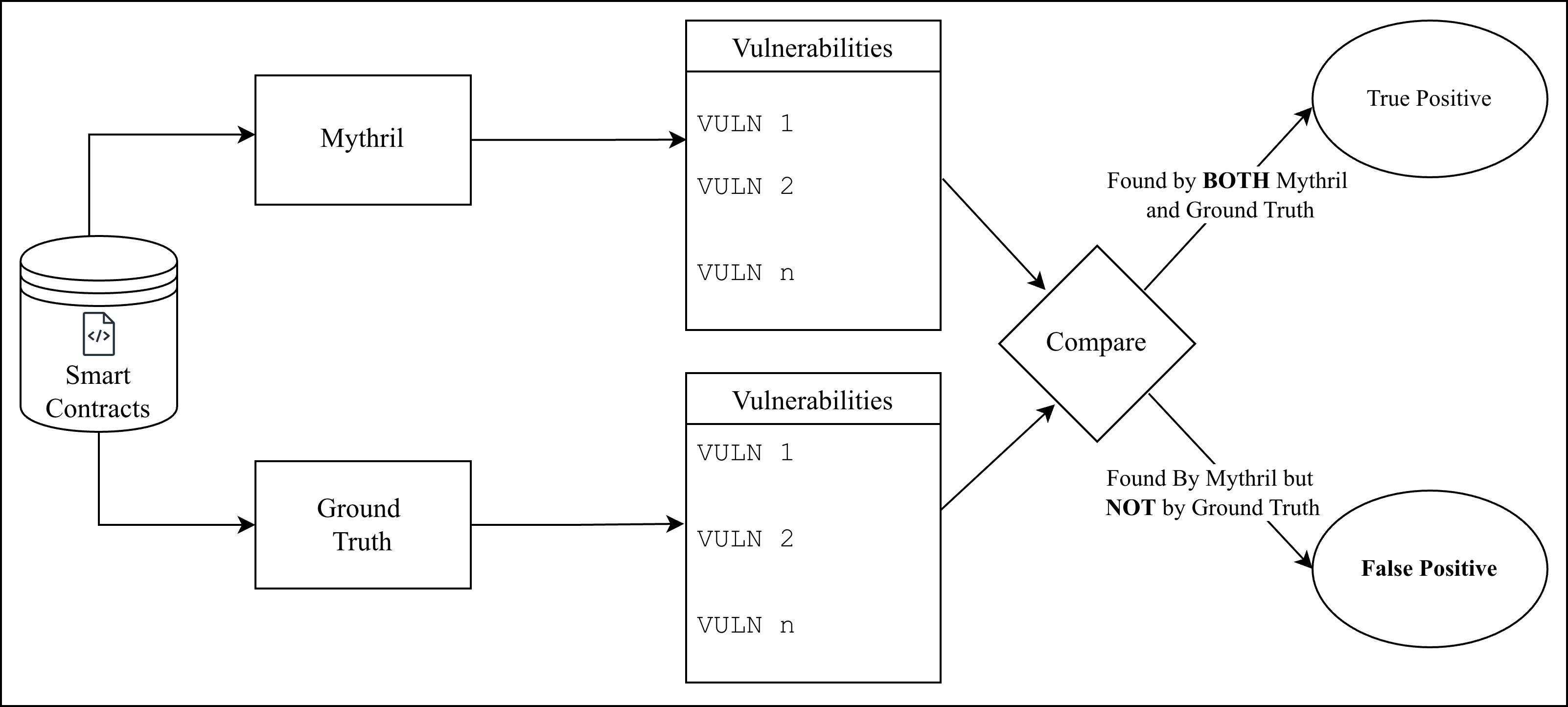}
  \caption{Process of analyzing Smart Contracts using Mythril and GigaHorse Benchmarks and comparing the reported vulnerabilities.}
  \label{fig:Process Diagram}
\end{minipage}\hfill % maximize horizontal separation
\end{figure}

We use vanilla Mythril \cite{a16} to analyze all Gigahorse Benchmarks \cite{a15} smart contracts. As expected, Mythril reported vulnerabilities in several contracts.

We utilized the labels provided within the Gigahorse Benchmarks as the ground truth to assess the accuracy of vulnerabilities reported by vanilla Mythril. This evaluation was performed on two distinct subsets:
\begin{itemize}
    \item \textbf{Vulnerable Contracts ($N=100$):} Contracts explicitly labeled with known vulnerabilities. If a vulnerability was reported by Mythril and matched the benchmark label, it was a \textbf{true positive (TP)}. If Mythril missed a known vulnerability, it was a \textbf{false negative (FN)}.
    \item \textbf{Invulnerable Contracts ($N=40$):} Contracts confirmed to be clean (benign). Any vulnerability flagged by Mythril in these contracts was classified as a \textbf{false positive (FP)}.
\end{itemize}

The overall evaluation process is illustrated in Fig.~\ref{fig:Process Diagram}. Our complete comparative analysis across the two Gigahorse subsets is presented in Table~\ref{tab:all-fp-results-combined}. A review of the results confirms that the Vanilla Mythril implementation generates a significant number of false positives across multiple vulnerability categories and datasets.

Specifically, in our baseline analysis of the Gigahorse Invulnerable contracts, vanilla Mythril reported a total of 102 false positives. In the next step, we incorporated the rules we proposed to reduce false positives into Mythril's implementation.

% Shadow old definitions and rebind to algpseudocode
\let\IF\relax \let\ENDIF\relax \let\ELSE\relax \let\ELSIF\relax
\let\FOR\relax \let\ENDFOR\relax \let\WHILE\relax \let\ENDWHILE\relax
\let\STATE\relax \let\RETURN\relax
\let\IF\If
\let\ENDIF\EndIf
\let\ELSE\Else
\let\ELSIF\ElsIf
\let\FOR\For
\let\ENDFOR\EndFor
\let\WHILE\While
\let\ENDWHILE\EndWhile
\let\STATE\State
\newcommand{\RETURN}[1]{\State \Return #1}
 
\section{Proposed Rules}
\label{sec:tool-improvement}

The following subsections discuss the implementation of proposed rules concerning six impactful vulnerabilities and how they improve the precision of the results produced by Mythril.

\subsection{State access after an external call}
\label{subsec: State access after an external call}
State access after an external call refers to the ability to read or modify the internal state of a smart contract after an external function call has been made. 
Previously, the \_check\_for\_reentrancy\_vulnerability method was called \textit{after} state changes in \_analyze\_state to detect potential re-entrancy vulnerabilities caused by those changes. 
This approach, however, led to false positives. 
To eliminate this, the code was modified to directly call the vulnerability check at every state change within \_analyze\_state, ensuring comprehensive evaluation and elimination of false positives.

The \_analyze\_state method in the StateChangeAfterCall module was modified to iterate over all StateChangeCallsAnnotation objects and their associated state\_change\_states. 
If a re-entrancy vulnerability is detected, the \_analyze\_state method creates a new PotentialIssue object and returns a list containing only that object. 
If no vulnerabilities are detected, the method returns an empty list. This ensures that only true positives are reported.

\begin{algorithm}
\caption{State access after an external call}
 \hrulefill
\begin{algorithmic}[1]
  \IF{length of offset $>$ 1}
    \FOR{ $i = 0$ to offset-1}
      \STATE offset to call 1
    \ENDFOR
  \ENDIF
  \IF{constraints == true}
    \RETURN false
  \ELSE
    \STATE new\_constraints = gas and destination address
    \IF{Generate transaction sequence is successful}
      \RETURN true
    \ENDIF
  \ENDIF
\end{algorithmic}
\label{algo:state-access}
\end{algorithm}

The \_check\_for\_reentrancy\_vulnerability first checks if any constraints in the state\_change\_state could prevent re-entrancy. 
Algorithm~\ref{algo:state-access} follows the following steps to correctly identify true positives only and not report any of false positives.

\begin{enumerate}
    \item It first checks if the current function being executed is the constructor, and if so, it returns an empty list.
    \item It then iterates over all StateChangeCallsAnnotation objects and their associated state\_change\_states.
    \item For each state\_change\_state, the method checks if it was preceded by a low-level call.
    \item If so, it creates a PotentialIssue object and adds it to the potential\_issues list of the StateChangeCallsAnnotation object.
    \item If there are constraints, it returns False since a re-entrancy vulnerability is not possible in that state.
    \item If there are no constraints, the method creates a new Constraints object and adds constraints to ensure that the gas and destination address of the external call meet certain conditions.
    \item It then tries to generate a transaction sequence from the global\_state to the change\_state.
    \item If the constraints are satisfied and a transaction sequence can be generated, it means that there is a potential re-entrancy vulnerability, and the method returns True.
\end{enumerate}

The \_analyze\_state method takes a GlobalState object as its argument and returns a list of PotentialIssue objects if re-entrancy vulnerabilities are detected.

\subsection{Assertion Violation}
\label{subsec: Assertion Violation}
An assertion violation occurs when a predefined condition within the implementation is not satisfied. It can occur when a condition or assumption results in false during execution. These assumptions or assertions are usually checks on the data validity, program state, or expected user input. An assertion violation is flagged when any of these checks are not satisfiable. It also signifies that the implementation is not prepared for unforeseen and unexpected situations. This could be due to a bug in the implementation, any unexpected user input, or a rare instance that was not handled in the implementation. 

Assertion violation belongs to the Exception class. The Exception class’s implementation is responsible for the detection of an assertion violation. The detection approach is followed by analyzing the entire execution state or the global state, and then identifying where the smart contract may reach an undefined state that contradicts an assertion. This specific approach is achieved by modifying hooks ($\_pre\_hooks$) within the smart contracts, these hooks serve as interception points or flags in the program’s execution flow. By strategically positioning these hooks, the Exceptions class gains deeper insight into program’s execution flow and can pinpoint potential assertion violations before they occur. Additionally, this strategic positioning allows additional control-flow decisions to accurately differentiate between true and false positives. This change within the implementation blocks the high number of false positives and reports the potential true positives before they occur. 

\begin{algorithm}[H]
 \hrulefill  \hrulefill
\caption{Assertion Violation}
\begin{algorithmic}[1]
\If {opcode is a CALL-related instruction}
    \State \textbf{Get} target address 
    \Comment{From The Stack}
    \If {it is an external call}
        \State \textbf{Add} instruction address to call offsets list
    \EndIf
\Else
    \If {there are multiple call offsets recorded}
        \For {each pair of consecutive call offsets}
            \State \textit{call\_1} $\gets$ first call offset
            \State \textit{call\_2} $\gets$ next call offset
            \State \textbf{Create} a list of constraints
            \State \textbf{Add} constraints 
            \Comment{from the WORLD STATE and mstate}
        \EndFor
    \EndIf
\EndIf
\end{algorithmic}
\end{algorithm}

The \_analyze\_state method performs the actual detection of assertion violations. The method first checks the current opcode to determine if it is a JUMP, INVALID, or REVERT opcode. If it is a JUMP opcode, the method annotates the current state with a LastJumpAnnotation object containing the address of the current instruction. If it is a REVERT opcode and not an assertion violation, the method simply returns an empty list. If it is an INVALID opcode and the previous opcode was a REVERT and the assertion condition is always false, the method considers this to be an assertion violation.  If an assertion violation is detected, the method creates an Issue object containing information about the violation and returns a list containing the issue. If no assertion violation is detected, the method returns an empty list.
The following changes were made to the code to achieve this:
    \begin{enumerate}
    \item The pre\_hooks of the detection module were changed to include "REVERT" to detect more assertion violations.
    \item The \_is\_assertion\_failure function was modified to check for both assertion violations and revert statements.
    \item The \_is\_assertion\_always\_false function was modified to check for both assertion failures and revert statements.
    \item The Exceptions class was modified to only return an issue if the current state is an assertion violation and the previous opcode is REVERT, indicating that the assertion condition is always false.
\end{enumerate}

If a vulnerability is detected, the algorithm creates an issue with relevant details and recommendations. Finally, it returns a list of identified issues. 
\subsection{Dependence on Predictable Environment Variable}
\label{subsec: Dependence on Predictable Environment Variable}
Dependence on predictable environment variables refers to a situation where the behavior or state of a smart contract relies on certain external factors that are expected to remain constant, such as external data, parameters, or conditions. If the external environment variables (i.e., parameters) were to change, this could result in false positives. We modified \_analyze\_state function where the control flow decisions were not accurate in the original code.

The modified \_analyze\_state function helps in achieving the objective of reducing false positive vulnerabilities and focusing on reporting the true positive ones. The function achieves this by the applying following rules and constraints: 
  
\begin{algorithm}[H]
\hrulefill
\hrulefill
\caption{Dependence on Predictable Environment Variable}
\begin{algorithmic}[1]
\If{the opcode is a CALL-related instruction}
    \State\textbf{Get} target address 
    \If{an external call}
        \State \textbf{Add} the address to the call offsets list
    \EndIf
\Else
    \If{multiple call offsets recorded}
        \For{each pair of consecutive call offsets}
            \State \textit{call\_1} $\gets$ First Call's Offset 
            \State \textit{call\_2} $\gets$ Next Call's Offset
            \State \textbf{Create} a list of constraints
        \EndFor
    \EndIf
\EndIf 
\end{algorithmic}
\label{algo: constraint dependency}
\end{algorithm}

\begin{enumerate}
    \item Checking if the control flow decision depends on one of the predictable environment variables (coinbase, gas limit, block number, or timestamp). 
    \item It monitors a tainted variable, which is a variable that can be modified by an outside user.
    \item Verifying whether the tainted value influences the constraints in the world\_state. 
    \item If it doesn't, the function considers this case as a false positive and doesn't report it.
    \item If the tainted value does influence the constraints, the function tries to find a valid transaction sequence that satisfies the constraints. 
    \item If a sequence is found, it generates an issue report for this vulnerability as a true positive.
\end{enumerate}

By following these steps, the updated function helps reduce the number of false positive vulnerabilities reported and focuses on reporting the true positive ones. This approach improves the accuracy of vulnerability detection and helps concentrate on true positive vulnerabilities.

\subsection{External Call to User-Supplied Address}
\label{subsec: External Call to User Supplied Address}
External Call to User Supplied Address refers to a scenario where a smart contract invokes a function or sends a transaction to an address provided as an argument or parameter by another contract. This allows smart contracts to interact with external contracts dynamically. The original code did not check for state modifications before and after the external calls. This often led to not identifying re-entrancy vulnerability. We changed the  \_analyze\_state function and added some constraints to solve this issue.

The \_analyze\_state function analyzes the given state, identifying potential issues related to insecure external calls in smart contracts. The proposed improvement helps in achieving the objective of reducing false positive vulnerabilities and focusing on reporting the true positives only. The function achieves this by the applying following rules and constraints:

This function, \_analyze\_state, analyzes the given state 
to identify potential issues related to insecure external calls in smart contracts.
\begin{algorithm}[H]
\hrulefill  \hrulefill
    \caption{External call to user-supplied address} 
\begin{algorithmic}[1]
    \If{tainted data = True} \Comment{Tainted data is present}
        \State constraintDependency $\gets$ [ ]
        
        \For{dependsOn in state.world.state.constraints} 
            \For{annotation in state.mstate.stack[-2].annotations}
                \If{annotation = predictable value annotation} 
                    \State constraintDependency $\gets$ constraintDependency $\cup$ \{annotation\}
                   \State \textbf{Add} annotation to constraintDependency Set
                \EndIf
            \EndFor
        \EndFor
        \State constraintDependency $\gets$ [ ]
    \EndIf 
\end{algorithmic}   
\end{algorithm}

To achieve the primary goal of reporting true positives while minimizing false positives, the function follows the following steps: 
\begin{enumerate}
    \item Ignore calls made in the constructor: The function starts by checking whether the current function is a constructor.
    \item If it is, the function returns an empty list, as it does not analyze constructor calls.
    \item Check for state modifications: The function inspects the instructions of the smart contract to identify state
    \item Modifications (SSTORE) both before and after the external call. This helps to determine if there is any state modification happening after the external call, which is a necessary condition for a reentrancy vulnerability to occur.
    \item Raise UnsatError for no state modifications after the call: If there are no state modifications after the external call, the function raises an UnsatError exception. 
    \item Add constraints for external calls: The function defines constraints to check for external calls with user-controlled addresses for any remaining gas. These constraints are necessary for identifying potential re-entrancy vulnerabilities.
    \item Find a valid transaction sequence: Using a solver, the function attempts to find a valid transaction sequence that satisfies the constraints. If the solver can find such a sequence, it indicates a potential re-entrancy issue, which is reported as a vulnerability.
\end{enumerate}
By following these steps,  the \_analyze\_state function focuses on reporting true positive vulnerabilities related to insecure external calls and reduces false positives by filtering out cases where the external call does not lead to a re-entrancy vulnerability. 

\subsection{Integer Arithmetic Overflow and Underflow}
\label{subsec: Integer Arithmetic Overflow}
Integer arithmetic overflow and underflow occur when the result of an arithmetic operation exceeds the maximum or falls below the minimum representable value for the data type being used. The current \_arithmetic\_helper function relied on a Z3 function and relied on the return value of the function to be true. This caused unpredictable behavior when the constraints in the \_arithmetic\_helper function were not precise enough. We modified this \_arithmetic\_helper function to perform checks on the Z3 solver itself.

The function bitvec\_helper uses Z3 \cite{a29}  to check whether a value is greater than or less than a certain threshold. The proposed improvement helps in achieving the objective of reducing false positive vulnerabilities and focusing on reporting the true positives only. The function achieves this by applying the following constraints and rules:
\begin{algorithm}
    \caption{Integer Arithmetic Overflow and Underflow}
    \hrulefill  \hrulefill
\begin{algorithmic}[1]
        \State maximum\_gas\_limit $\gets$ 1000000
        
        \State \textbf{Get} account balance: account\_balance       
        \If{$gas \leq \mathrm{MAX\_GAS\_LIMIT} \land gas \geq \mathrm{MIN\_GAS\_LIMIT}$}
            \If{$account\_balance > 0$}
                \State \textbf{return} satisfied \Comment{Constraints are satisfied}
            \Else
                \State \textbf{return} \textbf{!}satisfied \Comment{Account balance is too low.}
            \EndIf
        \Else
            \State \textbf{return} \textbf{!}satisfied \Comment {Gas limit exceeds the maximum.}
        \EndIf
\end{algorithmic}
\end{algorithm} 

Here's a breakdown of the steps: We create a constraint using the 'UGT' function from Z3, which stands for "Unsigned Greater Than". We propose to add the following rule to the solver, where we

\begin{enumerate}

    \item compares the value of 'raw' with a computed threshold value (2 * array. size() - 1) for overflow and (-2 * array. size() - 1) for underflow.
    \item We check if the constraints added to the solver are satisfiable by calling 'solver. check()'. If the result is 'sat' (satisfiable), it means there is a solution that satisfies the constraints. In this case, it indicates that the value of 'raw' is greater than the threshold, implying an overflow.
    \item After checking the satisfiability, we remove the added constraint from the solver. This step is essential only if we want to remove the constraints that were added during the solving process using backtracking.
\end{enumerate}

The \_arithmetic\_helper takes two objects and an operation. The objects must be of the BitVector type, and the operation must be specified using a callable object. The function performs the valid arithmetic operation on those objects and stores the annotations and the result in a new BitVector object. The result object has the same size as the input objects, and the annotations are the union of the annotations of the input objects. Then, the function checks for the potential overflow and underflow by comparing the result with the maximum and the minimum possible values of the input object's data type. The function returns the result object only if no overflow or underflow is detected. Applying these rules provide a convenient and safe way to perform arithmetic operations and reduce false positives.

\subsection{Multiple calls in a single transaction}
\label{subsec: Multiple calls in a single transaction}
Multiple calls in a single transaction refer to the ability to execute multiple function calls or interactions with different contracts within a single blockchain transaction. If a single function call is not successful, a false positive will be generated. The \_analyze\_state function did not check if the previous external call was successful or not. We modified the function and added some constraints to counter this and ensure that the reported vulnerabilities were genuine.

The proposed updates in the \_analyze\_state function reduce false positives by improving the checks and constraints for preventing multiple sends (i.e., external calls) within a single transaction. 

The additional checks and constraints ensure that the function only reports vulnerabilities when the previous external call was successful, thereby increasing the confidence that the reported vulnerabilities are genuine issues. 

\begin{algorithm}[H]
\caption{Multiple calls in a single transaction}
    \hrulefill 
\begin{algorithmic}[1]
 \State let x = state.world.state.constraints
 \State let y = state.mstate.constraints
 \State let z = state.mstate.stack
\If {len(call\_offsets) $>$ 1}
    \For{$i \gets (0$ to len(call\_offsets) - 1)}
        \State call\_1 $\gets$ call\_offsets[i] \Comment{First Call Offset}
        \State call\_2 $\gets$ call\_offsets[i + 1] \Comment{Next Call Offset}
        \State constraints $\gets$ x + [y[-1], y[-2]]
        \State constraints.append(z[-1] = 1)
    \EndFor
\EndIf
\end{algorithmic}
\end{algorithm}

Some of the improvements made to the function include: 
\begin{enumerate}
    \item Ensuring that at least two calls are present in the call\_offsets list before proceeding with the vulnerability detection logic.
    \item Verifying that the previous external call was successful by checking if the top element of the stack is equal to 1.
    \item Distinguish between external and internal calls by checking the target address of the calls.
\end{enumerate}
These improvements make the function more focused on true positive vulnerabilities and reduce the likelihood of false positives.

\section{Results and Evaluation} \label{resEva}

\begin{table}[t]
\caption{Summary of Performance Gains (Vanilla vs. Improved Mythril)}
\label{tab:summary-performance-metrics}
\centering
\renewcommand{\arraystretch}{1} % Increased slightly for better vertical spacing
\begin{tabular}{|l||c|c|}
\hline
\bfseries Metric & \bfseries Vanilla (V) & \bfseries Improved (I) \\
\hline
\hline
\multicolumn{3}{|l|}{\bfseries False Positive (FP) Reduction} \\
\hline
GH Invuln FP Total ($N=40$) & 102 $\to$ 51 & \textbf{50.0\% $\downarrow$} \\
GH Vuln FP Total ($N=100$) & 130 $\to$ 14 & \textbf{89.2\% $\downarrow$} \\
\hline
\multicolumn{3}{|l|}{\bfseries GH Vuln ($N=100$) Performance} \\
\hline
Precision ($\mathbf{TP} / (\mathbf{TP}+\mathbf{FP})$) & 8.5\% & \textbf{48.1\%} \\
FDR ($\mathbf{FP} / (\mathbf{TP}+\mathbf{FP})$) & \textbf{91.5\%} & \textbf{51.9\%} \\
Recall ($\mathbf{TP} / (\mathbf{TP}+\mathbf{FN})$) & 75.0\% & \textbf{81.2\%} \\
\hline
\textbf{F1-Score (Overall)} & \textbf{0.152} & \textbf{0.605} \\
\hline
\multicolumn{3}{p{0.25\textwidth}}{\footnotesize FDR: False Discovery Rate.} \\
\end{tabular}
\end{table}

\subsection{Gigahorse Benchmark Results}

We evaluate the efficacy of our proposed rule-based improvements by comparing Vanilla Mythril \cite{a16} against our Improved Mythril on the Gigahorse benchmark \cite{a15}, which consists of 40 invulnerable and 100 vulnerable smart contracts. Our core objective was to reduce the number of False Positives (FP) across the six targeted vulnerabilities. The complete findings are detailed in Table~\ref{tab:all-fp-results-combined}, with a summary of the performance metrics provided in Table~\ref{tab:summary-performance-metrics}.

\paragraph{False Positive Analysis}
The results show a significant reduction in false positives, verifying the tool's ability to perform accurate analysis:

\begin{itemize}
    \item \textbf{GigaHorse Invuln:} The Invuln or the invulnerable dataset is designed to accurately measure false positives FP. The total count of FPs decreased from $\mathbf{102}$ to $\mathbf{51}$, a $\mathbf{50.0\%}$ reduction. All FPs were eliminated with a $\mathbf{100.0\%}$ success rate for four (\emph{State Access after External Call}, \emph{Assertion Violation}, \emph{Dependence on Predictable Environment Variable}, and \emph{Multiple Calls in a Single Transaction}) of the six vulnerability types we implemented. 
    \item \textbf{GigaHorse Vuln:} The results on the Vuln or the vulnerable dataset are even more notable. The total count of FPs decreased from $\mathbf{130}$ to $\mathbf{14}$, a $\mathbf{89.2\%}$ reduction. This is a significant gain, as reflected by the False Discovery Rate (FDR), which decreased from $\mathbf{91.5\%}$ to $\mathbf{51.9\%}$.
\end{itemize}

% done till here

\paragraph{True Positives and Recall}
The proposed rules we implemented are designed to preserve and improve detection capabilities, enhance the precision of the tools, and correctly identify and report vulnerabilities. While the primary objective was FPs reduction, our results also demonstrate a positive shift in a tool’s ability to correctly identify and report true positives TPs. The total count of TPs increased from $\mathbf{12}$ to $\mathbf{13}$, which directly corresponds to the false negatives FNs that decreased from $\mathbf{4}$ to $\mathbf{3}$. This indicates that Mythril’s implementation with our proposed rules identified one vulnerability that the standard vanilla Mythril missed. Quantitatively, this small change translates to a rise in Recall from $\mathbf{75.0\%}$ to $\mathbf{81.2\%}$.

\subsection{Discussion: Overall Performance and Limitations}

We compare the standard Mythril (vanilla)  with the improved Mythril using standard performance metrics. Improved Mythril is the standard version of Mythril implemented using our proposed rules. We also report the F1-score, as it combines both precision and recall, and provides a comprehensive measure of overall accuracy. The improved Mythril reached an F1-score of $\mathbf{0.605}$, from $\mathbf{0.152}$ that vanilla Mythril achieved, reflecting a nearly $\mathbf{4\times}$ gain in overall performance. Table~\ref{tab:summary-performance-metrics} reports and compares the following performance metrics: Precision, False Discovery Rate, Recall, F1-Score, $\%$ FPs Reduction.

% The only notable outlier among the six targeted vulnerability classes is \emph{Integer Arithmetic: Overflow and Underflow}. This category contributes the majority of the remaining False Positives (12 out of 14) and all remaining False Negatives (3 out of 3). Furthermore, it was the only category where False Positives on the Gigahorse Invuln set increased (from 34 to 50), suggesting a need for more sophisticated rules specific to this complex class in future work. Despite this, the overall gains across the other five vulnerability types overwhelmingly validate the effectiveness of our proposed refinement strategy.

The only outlier among the six targeted vulnerability classes is \emph{Integer Arithmetic: Overflow and Underflow}, which contributes to the majority of the remaining FPs ($\mathbf{12}$ of $\mathbf{14}$) and all FNs ($\mathbf{3}$ of $\mathbf{3}$). In addition, it is the only vulnerability type for which the FPs increased from ($\mathbf{34}$ to $\mathbf{50}$) on the Invulnerable dataset. Further research is required to understand this unexpected behavior and then to develop more sophisticated rules specific to this complex class. Despite this, the substantial gains observed in Tables \ref{tab:all-fp-results-combined} and \ref{tab:summary-performance-metrics} for the other five most significant vulnerability types overwhelmingly validate the efficacy of our proposed rules.

\section{Methodology Generalization and Portability}
\label{sec:generalization}

We implemented our proposed rules in Mythril; however, the underlying methodology is designed for wide applicability. These rules are based on standard Ethereum Virtual Machine (EVM) principles and a small set of basic analysis primitives that are common to most analysis tools. Their design allows the logic to be easily separated and ported as a light layer to other symbolic analysis frameworks.

\subsection{Basic Analyzer Capabilities Required}
The minimum requirements for a target analyzer to successfully port proposed rules include four fundamental capabilities:

\begin{enumerate}
    \item \textbf{Path Feasibility Query:} The ability to determine if a specific execution path leading to a vulnerability alert can actually be reached by any valid transaction input. This typically uses an underlying constraint solver.
    \item \textbf{Data Taint Tracking:} The ability to mark input data (like transaction arguments or specific environment variables) and trace how that data influences the program's execution.
    \item \textbf{Call Type Classification:} The ability to accurately distinguish between internal jumps within the contract code and genuine external calls that cross address boundaries.
    \item \textbf{State Effect Detection:} The ability to identify when a specific execution path permanently alters the contract's storage or transfers assets.
\end{enumerate}

\subsection{Generalized Filtering Rules}
The primary goal of the rules is to suppress alerts where the detected vulnerability exists on a path that is infeasible or where a state check proves the alert is safe. The general principle applied to our six targeted vulnerability classes is as follows:

\begin{itemize}
    \item \textbf{Integer Arithmetic (Overflow/Underflow):} A wraparound is reported only if a path feasibility check confirms that inputs exist to trigger the overflow or underflow.
    \item \textbf{State Access after External Call (Re-entrancy):} Flag a state read or write made after an external call only when a feasibility check confirms no control flow guard (e.g., a re-entrancy lock) is active on that path.
    \item \textbf{Assertion Violation:} A program failure is flagged only if the failure condition depends on an external input that the path solver can prove to be always false or that leads to a feasible failure state.
    \item \textbf{Dependence on Predictable Environment Variable:} If a contract decision relies on a tainted, predictable environment value, an alert is issued only if a feasible path leads directly to a contract asset or storage change.
    \item \textbf{External Call to User-Supplied Address:} If an external call uses an address derived from tainted user input, the issue is reported only if a subsequent asset transfer or storage change is also proven feasible on that same path.
    \item \textbf{Multiple Calls in a Single Transaction:} Count the number of unique, true external message calls along a feasible path. The alert is triggered only if $\geq 2$ such calls are detected along with a resultant state effect.
\end{itemize}

\subsection{Portability and Limits}
\label{subsec:portability}
\textbf{Portability:} As the methodology relies only on these core principles, symbolic execution frameworks that expose similar capabilities, such as \emph{Oyente} \cite{a20} and \emph{Manticore} \cite{a17}, are conceptually suited for implementing this approach. This suggests that the measured precision gains are not exclusive to the tool used for implementation.

\textbf{Limits:} Our results are restricted to Mythril on the Gigahorse dataset (Section~\ref{resEva}), we do not claim performance for other tools. If a target tool lacks a required primitive (e.g., taint tracking), the corresponding filtering rule must be disabled or use a simplified check.

\section{Conclusion and Future Work}
\label{subsec: Conclusion and Future Work}

This paper addressed the issue of high False Positive (FP) rates in the symbolic analysis of Ethereum smart contracts by applying a set of lightweight, rule-based patches to Mythril across six key vulnerability classes. Using the Gigahorse benchmarks, our implementation achieved a significant reduction in False positives. Specifically, we reduced FPs by $\mathbf{50.0\%}$ on the invulnerable set and by $\mathbf{89.2\%}$ on the vulnerable set, leading to a near fourfold increase in the F1-Score, from a practically unusable $\mathbf{0.152}$ to $\mathbf{0.605}$ (Table~\ref{tab:summary-performance-metrics}). This improvement was achieved while also maintaining, and slightly improving, detection capability (Recall $75.0\% \to 81.2\%$).

\section{Future Work}
Future work may extend the scope and robustness of this study along three directions. First, in \emph{Integer Arithmetic}, investigate the unexpected behavior and develop more sophisticated rules that reduce the remaining false positives. Second, under \emph{Evaluation}, the ground truth reference may be expanded to include additional datasets for more rigorous testing of each proposed rule. Finally, in \emph{Portability}, the proposed rules are to be applied to the implementation of other tools to formally evaluate the generality of the approach.

\bibliographystyle{IEEEtran}
\bibliography{references}

\end{document}